# Time-dependent Bragg diffraction and short-pulse reflection by one-dimensional photonic crystals

Jean-Michel André*, Philippe Jonnard

*CNRS UMR 7614, Laboratoire de Chimie Physique - Matière et Rayonnement*
*Sorbonne Universités, UPMC Paris 06*
*11 rue Pierre et Marie Curie, F-75231 Paris Cedex 05, France*

*corresponding author : jean-michel.andre1@upmc.fr

**Abstract**

The time-dependence of the Bragg diffraction by one-dimensional photonic crystals and its influence on the short pulse reflection are studied in the framework of the coupled-wave theory. The indicial response of the photonic crystal is calculated and it appears that it presents a time-delay effect with a transient time conditioned by the extinction length. A numerical simulation is presented for a Bragg mirror in the x-ray domain and a pulse envelope modelled by a sine-squared shape. The potential consequences of the time-delay effect in time-dependent optics of short-pulses are emphasized.

Keywords: *time-dependent diffraction, short-pulse reflection, photonic crystal*

# 1. Introduction

Beside the particular physics of the Sommerfeld-Brillouin precursors [1–5], the study of the short pulse propagation in optical media has been mainly developed until now in the context of femtosecond lasers [6–8]. At the present time, there is a renewal of the optics of ultra-short pulses with the advent of femto-, subfemto- and attosecond sources in the high energy domain of the electromagnetic spectrum : attopulses are delivered by high harmonic generation (HHG) sources in the X-UV range [9–12], subfemto- and femtopulses are generated by X-ray free electron lasers (X-FEL) [13] or other new sources (Thomson scattering, …) in the x-ray domain [14,15]. These sources required optical devices (monochromators, mirrors, …) and the reflection properties of multilayer reflectors [16–18] and crystals [19,20] dedicated for this kind of sources have been treated by means of methods implemented for the frequency domain, the problem being that the pulse is a superposition of many harmonics which propagate with different velocities (wavelength dispersion) which results in pulse distortion.

The strategy for the calculation generally follows a Fourier mathematical approach. In this context, the Green impulse response corresponding to a Dirac-delta incident pulse is generally calculated since it allows calculating an arbitrary incident pulse response by means of the convolution theorem [19]. The design of chirped mirrors developed to compensate for the wavelength dispersion generally lies on the Fourier approach [20–24].

Diffraction of short pulses in the time domain is less investigated than in the frequency domain. The time-dependent propagation equation in crystals for the x-ray range (Tagaki-Taupin equation) was studied by Chukhovskii and Förster [25], then Wark and Lee [26] and more recently by Lindberg and Shvyd'ko [27]. In the present work, we adopt a rather similar approach for an one-dimensional photonic crystal (1D-PC) and we propose an analytical solution of the problem : the 1D-PC is modelled by a periodic stack of bilayers and the electric field of the optical pulse is written as a carrier wave modulated by a slow-varying envelope. Starting from Maxwell equations, the propagation time-dependent equation is deduced for the envelope and after some assumptions, a coupled system of partial differential equations (PDEs) is obtained for the two Fourier components of the two wave fields in the spirit of the coupled wave theory (CWT) [28]. This is done in Section 2. In Section 3, we present an analytic solution of the coupled system by using a matrix formalism. Then we calculate the reflectivity versus time for a constant intensity source abruptly switched on, or in other words, we determine the indicial response of the 1D-PC in Section 4. We show in Section 5, from a numerical example that the peak reflectivity versus time of the 1D-PC displays a transient period during which the reflectivity starts from a null value to reach the steady-state value with a rise-time determined by the extinction length; using Strejc's method, we then deduce from the computed indicial response, the transfer function which is the Laplacian transform of the impulse response. The knowledge of the transfer function gives access to the dynamics of reflection of a pulse of any shape by means of the convolution theorem ; the case of a sine-squared pulse is examined. In conclusion, Section 5, we discuss the implications of this time-delay effect, in particular for the x-ray regime and we give the perspectives.

**2. Time-dependent coupled-wave theory**

We consider the propagation of an optical pulse in the 1D-PC consisting in a periodic stack of bilayers as shown in Figure 1 ; the figure also gives the geometry of the problem and some notations. The axes are oriented so that the plane of incidence defined by the $z$-axis normal to the plane of stratification ($x$, $y$) and the optical axis defined by the incident wavevector $\boldsymbol{k} = (k_x, k_z)$ coincides with the ($x$, $z$) plane and we assume that the photonic crystal is uniform in the ($x$, $y$) plane so that we will neglect the trivial $y$ dependence. The angle $\theta$ is the glancing angle, that is the angle between the optical axis and the plane of stratification ($x$, $y$). In a general way, from macroscopic Maxwell's equations, it results that the electric field associated to the optical pulse satisfies the following wave equation written in the time domain using Gaussian units

$$\left(\frac{1}{c^2}\frac{\partial^2}{\partial t^2} - \nabla^2\right)\boldsymbol{E}(\boldsymbol{r},t) + \nabla(\nabla.\boldsymbol{E}(\boldsymbol{r},t)) = -\frac{4\pi}{c^2}\frac{\partial^2 \boldsymbol{P}(\boldsymbol{r},t)}{\partial t^2}$$

(1)

where $\boldsymbol{E}(\boldsymbol{r},t)$ stands for the electric field and $\boldsymbol{P}(\boldsymbol{r},t)$ for the electric polarization vector [29], $c$ being the speed of light in vacuum and $\boldsymbol{r} = (x, y, z)$ the position vector. Propagation takes place longitudinally, that is along the $z$-axis. In these conditions and within the paraxial approximation where the term $\nabla(\nabla.\boldsymbol{E}(\boldsymbol{r},t))$ is neglected, the wave equation reduces to

$$\frac{\partial^2 \boldsymbol{E}(x,z,t)}{\partial x^2} + \frac{\partial^2 \boldsymbol{E}(x,z,t)}{\partial z^2} - \frac{1}{c^2}\frac{\partial^2 \boldsymbol{E}(x,z,t)}{\partial t^2} = \frac{4\pi}{c^2}\frac{\partial^2 \boldsymbol{P}(x,z,t)}{\partial t^2}$$

$$\text{(2)}$$

The paraxial approximation is justified in the case of the propagation of transverse waves inside the 1D-PC.

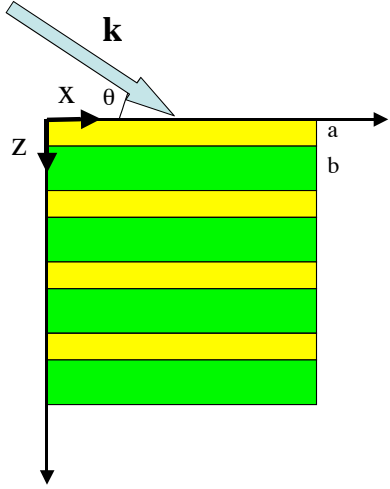

*Figure 1: Sketch of the 1D-PC consisting in a periodic stack of $\mathcal{N}$ bilayers with thickness d; the bilayer is made up of a material **a** with dielectric susceptibility $\chi_a$ and material **b** with dielectric susceptibility $\chi_b$ with layer thickness $d_a = \gamma d$ and $d_b = (1-\gamma)d$ respectively. The incoming radiation with a wavevector **k** in the plane (x, z) strikes the multilayer structure under a glancing angle $\theta$. A Cartesian orthogonal reference frame $(\hat{x}, \hat{y}, \hat{z})$ is used.*

We consider only the *s*-polarization (transverse electric) case so that the electric field vector $\boldsymbol{E}(x,z,t)$ is along the *y* axis. The *p*-polarisation (transverse magnetic) case is more complicated to treat because it involves the magnetic field with a supplementary term in the propagation equation difficult to handle; this case will be considered in a forthcoming work. It is assumed that the electric field of the optical pulse is formed of a quickly varying carrier with frequency $\omega$, modulated by an envelope $E_0(z,t)$ and we write it as follows taking into account the *s*-polarisation

$$\boldsymbol{E}(x,z,t) = E_0(z,t)\, e^{i(k_x x - \omega t)}\, \hat{\boldsymbol{y}}$$

$$\text{(3)}$$

A sine-squared limited in a time interval is usually adopted to model the envelope, especially for X-FEL single pulses [30] that is $E_0(z,t)$ varies as $\sin^2\left(\pi \frac{t}{T}\right)$ in the interval *[0,T]* and *0* outside. Let us emphasize that our approach does not depend on the pulse shape ; indeed we first search for the indicial response of the 1D-PC and the response to any physical pulse shape will be found from the indicial response using a method implementing the convolution theorem as explained in Section 4. We assume that the polarization is essentially electronic and follows instantly the change on the electric field, and that the media have a linear behaviour. Hence we write the polarisation **P** as

$$\boldsymbol{P}(x,z,t) = \chi(z)\,\boldsymbol{E}(z,t) = \chi(z)\,E_0(z,t)\,e^{i(k_x x - \omega t)}\,\hat{\boldsymbol{y}}$$

(4)

where $\chi(z)$ is the electric susceptibility assumed in our model to be time independent and varying only along the z direction. Inserting Eqs.(3,4) in Eq.(2) leads to the following equation for the propagation of the electric field envelope

$$\frac{\partial^2 E_0(z,t)}{\partial z^2} + k_\perp(z)^2 E_0(z,t) + 2i\frac{\omega}{c^2}\epsilon(z)\frac{\partial E_0(z,t)}{\partial t} - \frac{\epsilon(z)}{c^2}\frac{\partial^2 E_0(z,t)}{\partial t^2} = 0$$

(5)

with the dielectric constant $\epsilon$ given by

$$\epsilon(z) = 1 + 4\pi\chi(z)$$

(6)

and

$$k_\perp(z)^2 = \epsilon(z)\frac{\omega^2}{c^2} - k_x^2$$

(7)

At this stage, we precise the framework of our analysis; we adopt the slowly varying amplitude approximation (SVA) both in time and space : the second derivatives with respect to space and time are neglected, and the following conditions are fulfilled

$$\left|\frac{\partial \boldsymbol{E_0}(z,t)}{\partial z}\right| \ll \frac{\omega}{c}\,|\boldsymbol{E_0}(z,t)|$$

(8)

$$\left|\frac{\partial \boldsymbol{E_0}(z,t)}{\partial t}\right| \ll \omega\,|\boldsymbol{E_0}(z,t)|$$

(9)

The two SVA approximations are very different from each other ; the SVA in space holds if the variation of the electric field amplitude is small upon travelling a distance equal to the carrier wavelength, whilst the SVA in time is valid provided the electric field amplitude varies little within the carrier period at any position. As a result of the different approximations in particular the SVA in time, Eq.(5) can be reduced to a time-dependent Schrödinger equation

$$\frac{\partial^2 E_0(z,t)}{\partial z^2} + k_\perp(z)^2 E_0(z,t) + 2i\,\epsilon(z)\frac{\omega}{c^2}\frac{\partial E_0(z,t)}{\partial t} = 0$$

(10)

Note that at this step, the SVA in space has not yet been applied. Eqs.(9) could be treated by means of numerical codes developed to solve the time-dependent Schrödinger equation. We choose in this work to use instead the approach of CWT which proved to be very efficient in this kind of optical problem [31]. In a 1D-PC, the electric field can be written as the superposition of two waves propagating in opposite directions along the z-axis, so that we write in the spirit of the CWT

$$E_0(z,t) = F(z,t)\, e^{+i\kappa z} + B(z,t)\, e^{-i\kappa z}$$

(11a)

with

$$\kappa = k\, sin\theta$$

(11b)

The susceptibility in the 1D-PC can be expanded in Fourier series

$$\chi(z,t) = \bar{\chi} + \sum_{p=-\infty}^{+\infty} \Delta\chi\, u_p\, e^{i\, pg\, z}$$

(12a)

with

$$\bar{\chi} = \chi_a\, \gamma + \chi_b\, (1-\gamma)$$

(12b)

$$\Delta\chi = \chi_a - \chi_b$$

(12c)

$$u_p = \frac{-i}{2\, p\, \pi} (1 - e^{-2i\pi p\gamma})$$

(12d)

$$g = \frac{2\pi}{d}$$

(12e)

In order to get a system of differential equations with constant terms, it is convenient to introduce auxiliary amplitude terms

$$f(z,t) = F(z,t)\, \exp\left[-i\left(\frac{p\, g}{2} - \kappa\right) z\right]$$

(12a)

and

$$b(z,t) = B(z,t) \exp\left[+i\left(\frac{p\,g}{2} - \kappa\right)z\right]$$

(12b)

In the vicinity of the *pth* Bragg diffraction resonance $\frac{p\,g}{2} - \kappa \approx 0$ and $f(z,t) = F(z,t)$, $b(z,t) = B(z,t)$. Combining Eqs.(9-12) and performing the calculation in the framework of the assumptions given above, in particular using now the SVA in space, it comes after some algebra the following system of time-dependent coupled PDEs satisfied by the varying amplitudes $f(z,t)$ and $b(z,t)$ closed to the *pth* Bragg diffraction ; using the matrix formalism, this system reads

$$\frac{\partial}{\partial z}\overline{\mathcal{E}}(z,t) = \overline{T}\frac{\partial}{\partial t}\overline{\mathcal{E}}(z,t) + i\,\overline{\mathcal{M}}\,\overline{\mathcal{E}}(z,t)$$

(13a)

where $\overline{\mathcal{E}}(z,t)$ is the column amplitude vector

$$\overline{\mathcal{E}}(z,t) = \begin{pmatrix} f(z,t) \\ b(z,t) \end{pmatrix}$$

(13b)

$\overline{\mathcal{M}}$ is the propagation matrix in space

$$\overline{\mathcal{M}} = \begin{pmatrix} -\alpha & K^+ \\ K^- & \alpha \end{pmatrix}$$

(13c)

with

$$\alpha = \frac{p\,g}{2} + \frac{k^2}{\kappa} 2\pi\,\bar{\chi} - \kappa$$

(13d)

$$K^+ = -\frac{k^2}{\kappa} 2\pi\,\Delta\chi\,u_p$$

(13c)

$$K^- = \frac{k^2}{\kappa} 2\pi\,\Delta\chi\,u_{-p}$$

(13d)

$\overline{T}$ is the propagation matrix in time

$$\overline{T} = \begin{pmatrix} -\frac{k}{c\,\kappa} & 0 \\ 0 & \frac{k}{c\,\kappa} \end{pmatrix}$$

(13e)

## 3. Matrix formalism for the CWT

We first consider the time-independent case

$$\frac{\partial}{\partial z} \overline{\mathcal{E}}(z) = i \, \overline{\mathcal{M}} \, \overline{\mathcal{E}}(z)$$

(14)

The solution can be obtained by substituting

$$\overline{\mathcal{E}}(z) = \begin{pmatrix} A \\ B \end{pmatrix} e^{i \psi z}$$

(15)

and as shown in Appendix I,

$$\overline{\mathcal{E}}(z) = \bar{S}(z) \, \overline{\mathcal{E}}(0)$$

(16)

with

$$\bar{S}(z) = \begin{pmatrix} \cos qz - i \frac{\alpha}{q} \sin qz & i \frac{K^+}{q} \sin qz \\ i \frac{K^-}{q} \sin qz & \cos qz + i \frac{\alpha}{q} \sin qz \end{pmatrix}$$

(17a)

with

$$q = \sqrt{K^+ K^- + \alpha^2}$$

(17b)

Returning to the time-dependent case, one searches the solution by analogy with the time-independent case in the following form :

$$\overline{\mathcal{E}}(z,t) = \begin{pmatrix} A(t) \\ B(t) \end{pmatrix} e^{i \psi z}$$

(18)

Inserting Eq.(18) in Eq.(13a) gives after derivation with respect to space

$$\frac{\partial}{\partial t} \begin{pmatrix} A(t) \\ B(t) \end{pmatrix} = -\bar{G} \begin{pmatrix} A(t) \\ B(t) \end{pmatrix}$$

(19)

with

$$\bar{G} = i \, \bar{T}^{-1}(\psi \bar{I} - \overline{\mathcal{M}})$$

(20)

Integration on time gives

$$\begin{pmatrix} A(t) \\ B(t) \end{pmatrix} = \exp(-\bar{G} \, t) \begin{pmatrix} A(0) \\ B(0) \end{pmatrix}$$

(21)

Finally it comes by following a way similar to the one presented for the time-independent case

$$\overline{\mathcal{E}}(z,t) = R(z,t)\ \overline{\mathcal{E}}(0,0)$$

(22)

where

$$R(z,t) = \exp(-\bar{G}\ t)\ \bar{S}(z)$$

(23)

From Eqs.(22,23) it is possible to calculate the distribution of the electric field in space and time, and also the reflection and the transmission coefficients. In the next section, we focus on the reflection coefficient under Heaviside unit-step input to obtain the indicial response.

## 4. Indicial response

The reflection coefficient follows from the initial and boundary conditions : at $z = 0$, a Heaviside unit-step $\Theta(t)$ is applied, so that $f(0,t) = \Theta(t)$, and at $z = L$ there is no incoming wave, so that $b(L,t) = 0$, which gives

$$\begin{pmatrix} f(L,t) \\ b(L,t) = 0 \end{pmatrix} = \begin{pmatrix} R_{11}(L,t) & R_{12}(L,t) \\ R_{21}(L,t) & R_{22}(L,t) \end{pmatrix} \begin{pmatrix} \Theta(0^+) \\ b(0,0^+) \end{pmatrix}$$

(24)

Then

$$b(0,0^+) = -\frac{R_{21}(L,t)}{R_{22}(L,t)}$$

(25)

To calculate the reflection coefficient, we need $b(0,t)$. From Eq.(25)

$$\begin{pmatrix} f(0,t) \\ b(0,t) \end{pmatrix} = \bar{R}(0,t) \begin{pmatrix} f(0,0^+) = \Theta(0^+)) \\ b(0,0^+) \end{pmatrix}$$

(26)

or

$$b(0,t) = R_{21}(0,t) + R_{22}(0,t)\ b(0,0^+)$$

(27)

Combining Eqs.(25) and (27) gives

$$b(0,t) = R_{21}(0,t) - R_{22}(0,t)\ \frac{R_{21}(L,t)}{R_{22}(L,t)}$$

(28)

Consequently using the definition of the reflection coefficient, one finds the indicial response $R_\Theta$ in terms of reflectivity at the time $t$ after switching on abruptly a constant intensity source at $t = 0$, or in other words, when applying a Heaviside unit-step input $\Theta$; it comes for the indicial response

$$R_\Theta(t) \equiv \left|b(0,t)\right]_0^t\right|^2$$

(29)

## 5. Numerical simulations

Since development of short pulse sources is rapidly growing in the high energy range of the electromagnetic spectrum, we choose to carry out our simulation in the x-ray domain for a 1D-PC consisting in a Bragg reflector formed by a Fe/C periodic stack of $\mathcal{N}$ bilayers; the period $d$ is equal to 5.0 nm and the $\gamma$ ratio is equal to 0.5, that is to say the thicknesses of the Fe and C layers are the same. The energy of the incident radiation is 8 keV, which gives a Bragg angle of 0.93°. Figure 2 displays the steady-state reflectivity versus $\theta$ angle for $\mathcal{N} = 100$; the value of the peak reflectivity (PR) is around 0.64. In all calculations we use for the optical indices, the values tabulated in the CXRO database [32].

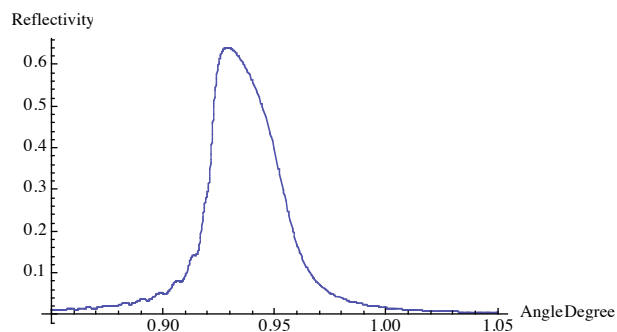

*Figure 2: Steady-state reflectivity versus the glancing angle $\theta$ at 8 keV of the 1D-PC, the parameters of which are given in the main text.*

Figure 3 shows the instantaneous PR, hereafter denoted $\hat{R}_\Theta(t)$, calculated at 8 keV for a constant intensity source switched on abruptly at $t = 0$; this quantity is the indicial response in terms of peak reflectivity. Let us note that the superscript $\widehat{\phantom{x}}$ will be used to denote the PR obtained at the Bragg angle which must be distinguished from the reflectivity without superscript obtained at any glancing angle $\theta$. One observes that $\hat{R}_\Theta(t)$ starts from zero then grows and tends asymptotically to the steady-state value (0.64) after a transient period of duration $\tau_t$ estimated to be around 42 fs. We define the transient time $\tau_t$ as the rise-time required to go from 10% to 90% of the asymptotic PR. For the 1D-PC and in the *s*-polarisation case, the extinction length $L_{exc}$ is given by the formula [27]

$$L_{exc} = \frac{\lambda \sin\theta}{4\pi |Re(\Delta\chi)| |u_p|}$$

(30)

In our example, $L_{exc}$ takes a value close to 325 nm, so that the characteristic time $\tau_c = L_{exc}/(\pi \sin\theta\, c)$ is around 20 fs. The time-dependent diffraction in crystals [25] shows that the saturation time $\tau_s$ is of the order of a few $\tau_c$: $\tau_s = \sigma\, \tau_c$ with $\sigma$ around 2.1. The latter formula gives 42 fs for $\tau_s$ in agreement with our transient time value $\tau_t$.

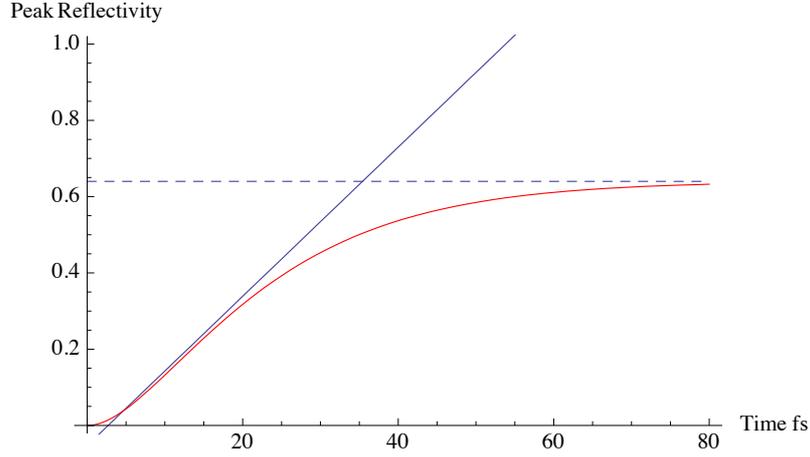

Figure 3: Instantaneous peak reflectivity $\hat{R}_\Theta(t)$ at 8 keV of the Fe/C 1D-PC, the parameters of which are given in the main text (red curve), steady-state peak reflectivity (dashed blue curve), tangent line at the inflection point of the instantaneous reflectivity (solid blue line).

Let us now consider the transfer function $\hat{R}_\delta(s)$ in terms of peak reflectivity that is the Laplacian transform of the impulse response $\hat{R}_\delta(t)$. This impulse response corresponds to the instantaneous peak reflectivity obtained when the system is struck by a Dirac pulse at *t=0* and must not be confused with the indicial response $\hat{R}_\Theta(t)$. There are a lot of methods to deduce the transfer function $\hat{R}_\delta(s)$ from the indicial response $\hat{R}_\Theta(t)$. In a first basic approach, we apply the graphic method introduced by Strejc [33]. The transfer function $\hat{R}_\delta(s)$ is written on the following form for an aperiodic system without delay :

$$\hat{R}_\delta(s) = \frac{K}{(1 + \tau s)^n}$$

(32)

where $K$, $n$ and $\tau$ are parameters to be determined from the curve $\hat{R}_\Theta(t)$; details of the method can be found in [34]. In our example, one finds $n = 1$ and $\tau = 35.34$ fs so that

$$\hat{R}_\delta(s) = \frac{K}{1 + \tau s} \approx \frac{0.64}{1 + 35.34\, s}$$

(33)

For an incident pulse with envelope shape $E(t)$, then by virtue of the convolution theorem, the Laplace transform $S(s)$ of the time-depend reflected pulse $S(t)$ is related to the Laplace transform $E(s)$ of $E(t)$:

$$S(s) = E(s)\, \hat{R}_\delta(s)$$

(34)

If $E(t)$ has a sine-squared shape ($\sin^2\left(\pi \frac{t}{T}\right)$ in the interval [0,T] and 0 outside), then the Laplacian transform $S(s)$ reads

$$S(s) = \frac{2\, K\, \pi^2\, (1 - e^{-sT})\Theta(T)}{(1 + s\, \tau)(4\, \pi^2\, s + s^3\, T^2)}$$

(35)

where $\Theta(T)$ stands for the Heaviside unit-step function. The formula for the inverse Laplacian transform of $S(s)$ is too long to be explicitly given in the text, but Figure 4 displays $S(t)$ for an incident pulse of width $T$ = 10 fs, for the system under consideration.

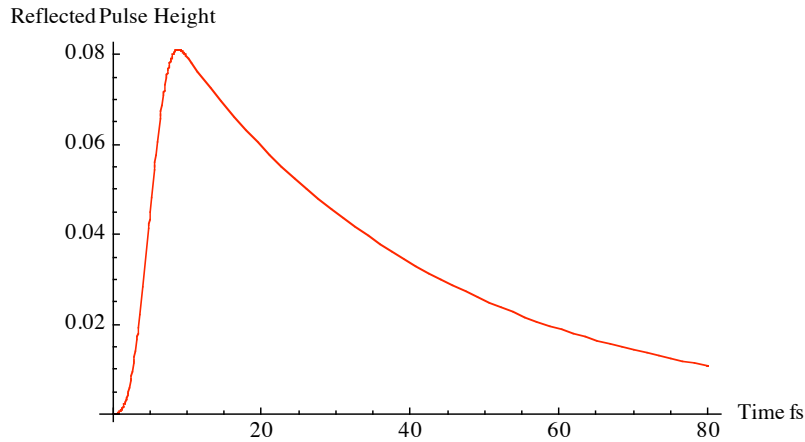

Figure 4: *Time dependence of the reflected pulse height S(t) with a sine-squared incident pulse of width T = 10 fs for the system under consideration.*

One observes clearly the time-delay effect : at $t$ = 10 fs corresponding to the width of the incident pulse, the reflected pulse reaches its maximum intensity which is only 8%, about 1/8 of the steady-state value; let us also note that the full width at half maximum is about 25 fs, much larger than the initial pulse width.

Finally let us consider briefly the influence of different parameters ; first the number of bilayers $\mathcal{N}$: it determines the steady-state value of the peak reflectivity but does not substantially change the saturation time. On the contrary, the nature of the materials influences both peak reflectivity and saturation time as shown in Figure 5 ; this figure displays the instantaneous peak reflectivity $\hat{R}_\Theta(t)$ at 8 keV for $\mathcal{N}$=100 but with the iron (Fe) layer replaced by a tungsten (W) or molybdenum (Mo) layer. The material dependence on $\hat{R}_\Theta(t)$ arises through the dielectric constant, which affects both peak reflectivity and extinction length.

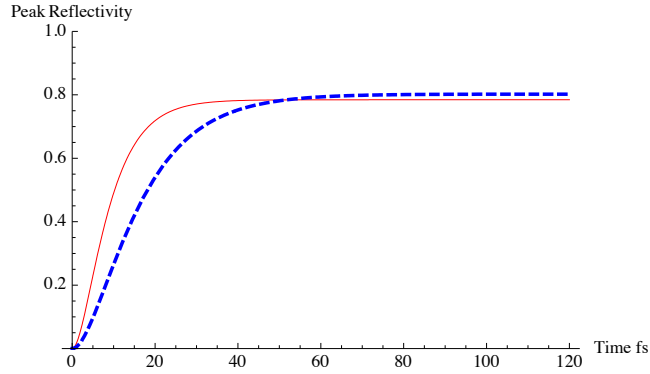

Figure 5: Instantaneous peak reflectivity $\hat{R}_\Theta(t)$ at 8 keV of the W/C 1D-PC (red solid line) and the Mo/C 1D-PC (blue dashed line) with the same parameters than the Fe/C 1D-PC.

## 6. Conclusion

The coupled-wave theory has been successfully implemented to treat the problem of time-dependent Bragg reflection by a 1D-PC. In the framework of this approach, we have shown that the instantaneous peak reflectivity of 1D-PC under unit-step striking presents a time-delay effect : this reflectivity starts from a zero value at the switching-on time and tends asymptotically towards the steady-state value. The transient time is conditioned by the extinction length. This result is in agreement with the conclusions drawn from the study of the time-dependent diffraction by crystals [25–27]. Our conclusion has potential consequences in time-dependent optics, in particular x-ray optics because x-ray pulse sources such as X-FEL have time duration comparable with the transient time of the multilayer Bragg mirrors.

An important but no trivial extension of the present study is to incorporate a pertinent description of the possible distortion of the structure due to the large power which can be deposited in the materials. Indeed short pulses sources, because they generally deliver high power (large energy in short duration), can distort or even damage the optics [35,36].

**APPENDIX I**

Inserting Eq.(15) in Eq.(14) leads to eigenvalue problem

$$(\overline{\mathcal{M}} - \psi \bar{I}) \begin{pmatrix} A \\ B \end{pmatrix} = 0$$

(A.1)

Solving this problem gives the eigenvalues $\psi^\pm$ of $\overline{\mathcal{M}}$

$$\psi^\pm \equiv \pm q \; ; q = \sqrt{K^+ K^- + \alpha^2}$$

(A.2)

To the two eigenvalues $\psi^\pm$ are associated two eigenvectors $\bar{V}^\pm$

$$\bar{V}^\pm = \begin{pmatrix} K^+ \\ \pm q \end{pmatrix}$$

(A.3)

The solutions for $\overline{\mathcal{E}}(z)$ can be derived using the eigenmatrix

$$\bar{P} = (\bar{V}^+ \quad \bar{V}^-) = \begin{pmatrix} K^+ & K^+ \\ +q - \alpha & -q - \alpha \end{pmatrix}$$

(A.4)

that is

$$\overline{\mathcal{E}}(z) = \bar{P} \begin{pmatrix} e^{+iqz} & 0 \\ 0 & e^{-iqz} \end{pmatrix} \bar{P}^{-1} \overline{\mathcal{E}}(0)$$

Since the general solution is a linear combination of the eigensolutions

$$\overline{\mathcal{E}}(z) = C^+ e^{+iqz} \bar{V}^+ + C^- e^{-iqz} \bar{V}^- = \bar{P} \begin{pmatrix} e^{+iqz} & 0 \\ 0 & e^{-iqz} \end{pmatrix} \begin{pmatrix} C^+ \\ C^- \end{pmatrix}$$

(A.5)

At z = 0, one has

$$\begin{pmatrix} C^+ \\ C^- \end{pmatrix} = \bar{P}^{-1} \overline{\mathcal{E}}(0)$$

(A.6)

Putting Eq.(A.6) in Eq (A.5), it follows that

$$\overline{\mathcal{E}}(z) = \bar{S}(z) \overline{\mathcal{E}}(0)$$

(A.7)

where $\bar{S}(z)$ is obtained by the product $\bar{P} \begin{pmatrix} e^{+iqz} & 0 \\ 0 & e^{-iqz} \end{pmatrix} \bar{P}^{-1}$

References

[1]     Sommerfeld A 1914 Über die Fortpflanzung des Lichtes in dispergierenden Medien *Ann. Phys.* **349** 177–202
[2]     Brillouin L 1914 Über die Fortpflanzung des Lichtes in dispergierenden Medien *Ann. Phys.* **349** 203–40
[3]     Oughstun K and Sherman S 1994 *Electromagnetic Pulse Propagation in Casual Dielectrics* (Springer-Verlag)
[4]     Uitham R and Hoenders B J 2006 The Sommerfeld precursor in photonic crystals *Optics Communications* **262** 211–9
[5]     Uitham R and Hoenders B J 2008 The electromagnetic Brillouin precursor in one-dimensional photonic crystals *Optics Communications* **281** 5910–8
[6]     Akhmanov S A, Vysloukh V A and Chirkin A S 1992 *Optics of Femtosecond Laser Pulses* (New-York: American Institute of Physics)
[7]     Rulliere C 2005 *Femtosecond Laser Pulses - Principles and Experiments* (Springer-Verlag)
[8]     Keller U 2010 Ultrafast solid-state laser oscillators: a success story for the last 20 years with no end in sight *Appl. Phys. B* **100** 15–28
[9]     Sansone G, Benedetti E, Calegari F, Vozzi C, Avaldi L, Flammini R, Poletto L, Villoresi P, Altucci C, Velotta R, Stagira S, Silvestri S D and Nisoli M 2006 Isolated Single-Cycle Attosecond Pulses *Science* **314** 443–6
[10]    Chen M-C, Mancuso C, Hernández-García C, Dollar F, Galloway B, Popmintchev D, Huang P-C, Walker B, Plaja L, Jaroń-Becker A A, Becker A, Murnane M M, Kapteyn H C and Popmintchev T 2014 Generation of bright isolated attosecond soft X-ray pulses driven by multicycle midinfrared lasers *PNAS* **111** E2361–7
[11]    Paul P M, Toma E S, Breger P, Mullot G, Augé F, Balcou P, Muller H G and Agostini P 2001 Observation of a Train of Attosecond Pulses from High Harmonic Generation *Science* **292** 1689–92
[12]    Ko D H, Kim K T, Park J, Lee J and Nam C H 2010 Attosecond chirp compensation over broadband high-order harmonics to generate near transform-limited 63 as pulses *New J. Phys.* **12** 063008
[13]    Feldhaus J, Arthur J and Hastings J B 2005 X-ray free-electron lasers *J. Phys. B: At. Mol. Opt. Phys.* **38** S799
[14]    Schoenlein R W, Leemans W P, Chin A H, Volfbeyn P, Glover T E, Balling P, Zolotorev M, Kim K-J, Chattopadhyay S and Shank C V 1996 Femtosecond X-ray Pulses at 0.4 Å Generated by 90° Thompson Scattering: A Tool for Probing the Structural Dynamics of Materials *Science* **274** 236–8
[15]    André J-M, Le Guen K and Jonnard P 2014 Feasibility considerations of a soft-x-ray distributed feedback laser pumped by an x-ray free electron laser *Laser Phys.* **24** 085001
[16]    Ksenzov D, Grigorian S and Pietsch U 2008 Time–space transformation of femtosecond free-electron laser pulses by periodical multilayers *Journal of Synchrotron Radiation* **15** 19–25
[17]    Ksenzov D, Grigorian S, Hendel S, Bienert F, Sacher M D, Heinzmann U and Pietsch U 2009 Reflection of femtosecond pulses from soft X-ray free-electron laser by periodical multilayers *phys. stat. sol. (a)* **206** 1875–9
[18]    Bushuev V and Samoylova L 2011 Application of quasi-forbidden multilayer Bragg reflection for monochromatization of hard X-ray FEL SASE pulses *Nuclear Instruments and Methods in Physics Research Section A: Accelerators, Spectrometers, Detectors and Associated Equipment* **635** S19–23

[19]  Shastri S D, Zambianchi P and Mills D M 2001 Dynamical diffraction of ultrashort X-ray free-electron laser pulses *Journal of Synchrotron Radiation* **8** 1131–5

[20]  Bushuev V A 2008 Diffraction of X-ray free-electron laser femtosecond pulses on single crystals in the Bragg and Laue geometry *Journal of Synchrotron Radiation* **15** 495–505

[21]  Szipöcs R, Spielmann C, Krausz F and Ferencz K 1994 Chirped multilayer coatings for broadband dispersion control in femtosecond lasers *Opt. Lett.* **19** 201–3

[22]  Szipöcs R and Kőházi-Kis A 1997 Theory and design of chirped dielectric laser mirrors *Appl Phys B* **65** 115–35

[23]  Beigman I L, Pirozhkov A S and Ragozin E N 2002 Reflection of few-cycle x-ray pulses by aperiodic multilayer structures *J. Opt. A: Pure Appl. Opt.* **4** 433

[24]  Bourassin-Bouchet C, Diveki Z, de Rossi S, English E, Meltchakov E, Gobert O, Guénot D, Carré B, Delmotte F, Salières P and Ruchon T 2011 Control of the attosecond synchronization of XUV radiation with phase-optimized mirrors *Opt. Express* **19** 3809–17

[25]  Chukhovskii F N and Förster E 1995 Time-dependent X-ray Bragg diffraction *Acta Crystallographica Section A Foundations of Crystallography* **51** 668–72

[26]  Wark J S and Lee R W 1999 Simulations of femtosecond X-ray diffraction from unperturbed and rapidly heated single crystals *Journal of Applied Crystallography* **32** 692–703

[27]  Lindberg R R and Shvyd'ko Y V 2012 Time dependence of Bragg forward scattering and self-seeding of hard x-ray free-electron lasers *Phys. Rev. ST Accel. Beams* **15** 050706

[28]  Kogelnik H 1969 Coupled wave theory for thick hologram gratings *Bell System Tech. J.* **48** 2909–47

[29]  Jackson J D 1975 *Classical Electrodynamics* (Wiley)

[30]  Lan K, Fill E E and Meyer-ter-Vehn J 2003 Simulation of He-α and Ly-α soft X-ray lasers in helium pumped by DESY/XFEL-radiation *EPL* **64** 454

[31]  Yeh P 2005 *Optical Waves in Layered Media* (Wiley)

[32]  CXRO X-Ray Interactions With Matter *http://henke.lbl.gov/optical_constants/*

[33]  Strejc V 1959 Approximation aperiodisscher Ubertragungscharakteristiken *Regelungstechnik* **7** 124–8

[34]  De Larminat P 2007 *Analysis and Control of Linear Systems* (London: P. de Larminat)

[35]  Hau-Riege S P, London R A, Chapman H N and Bergh M 2007 Soft-x-ray free-electron-laser interaction with materials *Phys. Rev. E* **76** 046403

[36]  Peyrusse O 2012 Coupling of detailed configuration kinetics and hydrodynamics in materials submitted to x-ray free-electron-laser irradiation *Phys. Rev. E* **86** 036403